\documentclass[letterpaper, 10 pt, conference]{ieeeconf}  

\IEEEoverridecommandlockouts                              

\title{\LARGE \bf
The Individual-Targeting Assumption: A Systematic Review of Proactive Robots in Human Group Settings  
}

\author{Tauhid Tanjim$^{1}$, Tasmia Mayen $^{2}$, Malte F. Jung$^{1}$, and Susan R. Fussell$^{1}$
\thanks{*This work was not supported by any organization}
\thanks{$^{1}$T. Tanjim, M. F. Jung, S. R. Fussell are with the Department of Information Science at Cornell University, Ithaca, NY 14850, USA
        {\tt\small \{tt485, mfj28, srf72\}@cornell.edu}
        }%
\thanks{$^{2}$T. Mayen is with the School of Education, Louisiana State University,
        Baton Rouge, LA 70802, USA
        {\tt\small tmayen2@lsu.edu}}%
}
\usepackage{graphicx}
\usepackage{booktabs, tabularx, multirow, array, xcolor, colortbl, caption, makecell}
\usepackage{url}
\usepackage{hyperref}
\usepackage{placeins}  
\usepackage{float}      
\definecolor{headerblue}{RGB}{220,230,242}
\definecolor{lightgray}{RGB}{245,245,245}
\newcolumntype{L}[1]{>{\raggedright\arraybackslash}p{#1}}
\newcolumntype{C}[1]{>{\centering\arraybackslash}p{#1}}
\usepackage[table]{xcolor}
\usepackage{colortbl}
\usepackage{booktabs}
\usepackage{tabularx}
\usepackage{array}
\usepackage{makecell}
\definecolor{hdrblue}{RGB}{214,227,243}
\definecolor{rowgray}{RGB}{245,245,245}
\PassOptionsToPackage{table}{xcolor}
\usepackage{caption}
\begin{document}
\maketitle
\thispagestyle{empty}
\pagestyle{empty}

\begin{abstract}
Proactive robots are increasingly deployed in public environments where people are encountered not as isolated individuals but as members of cohesive social groups. 
Yet whether the prevailing design paradigm in proactive human-robot interaction (HRI) accounts for the relational structure that defines a group as a social unit remains largely unexamined.
Through a systematic review of 63 proactive HRI studies in group settings from 2000 to 2025, we identify a recurring tendency, the Individual-Targeting Assumption (ITA), in which robots treat co-present people as independent engagement targets. 
We find that ITA is present in 60.3\% of the corpus, with group-aware approaches emerging almost entirely after the robot is already embedded in an ongoing interaction.
Critically, how a robot should detect and negotiate entry into a pre-formed group before initiating contact remains unaddressed across the corpus.
Three failure modes, \textit{engagement misdetection}, \textit{social ratification blindness}, and \textit{bystander neglect}, emerge as recurring patterns when proactive robots interact with human groups.
These findings reframe proactive HRI in group settings not as an extension of dyadic interaction but as a qualitatively distinct design problem, and they identify the robot's approach to the group during the entry phase as a critical and understudied open challenge.




\end{abstract}
 
\section{Introduction}
The use of robots is expanding and becoming increasingly common in the public sphere, including shopping malls~\cite{kanda2010communication}, hospitals~\cite{tanjim2025human, tanjim2025help, taylor2024towards}, and schools~\cite{pusceddu2025will}. 
People are rarely isolated in such settings, as they arrive together, move together, and attend to their surroundings as socially coordinated units.
Robot interactions with a human group are therefore more complex than one-on-one interactions; in these public places, simultaneous interaction with multiple people is not the exception but the norm~\cite{sebo2020robots}.
Robots must thus be able to interpret, understand, navigate, and respond appropriately to human groups~\cite{gomez2025developing}, whose social structures, shared attention, and interpersonal dynamics are further shaped by the robot's presence~\cite{sebo2020robots}. 
Robots' seamless interaction with human groups has become an increasingly important research priority~\cite{gomez2025developing}.
\begin{figure}[!t]
    \centering
    \includegraphics[width=\columnwidth]{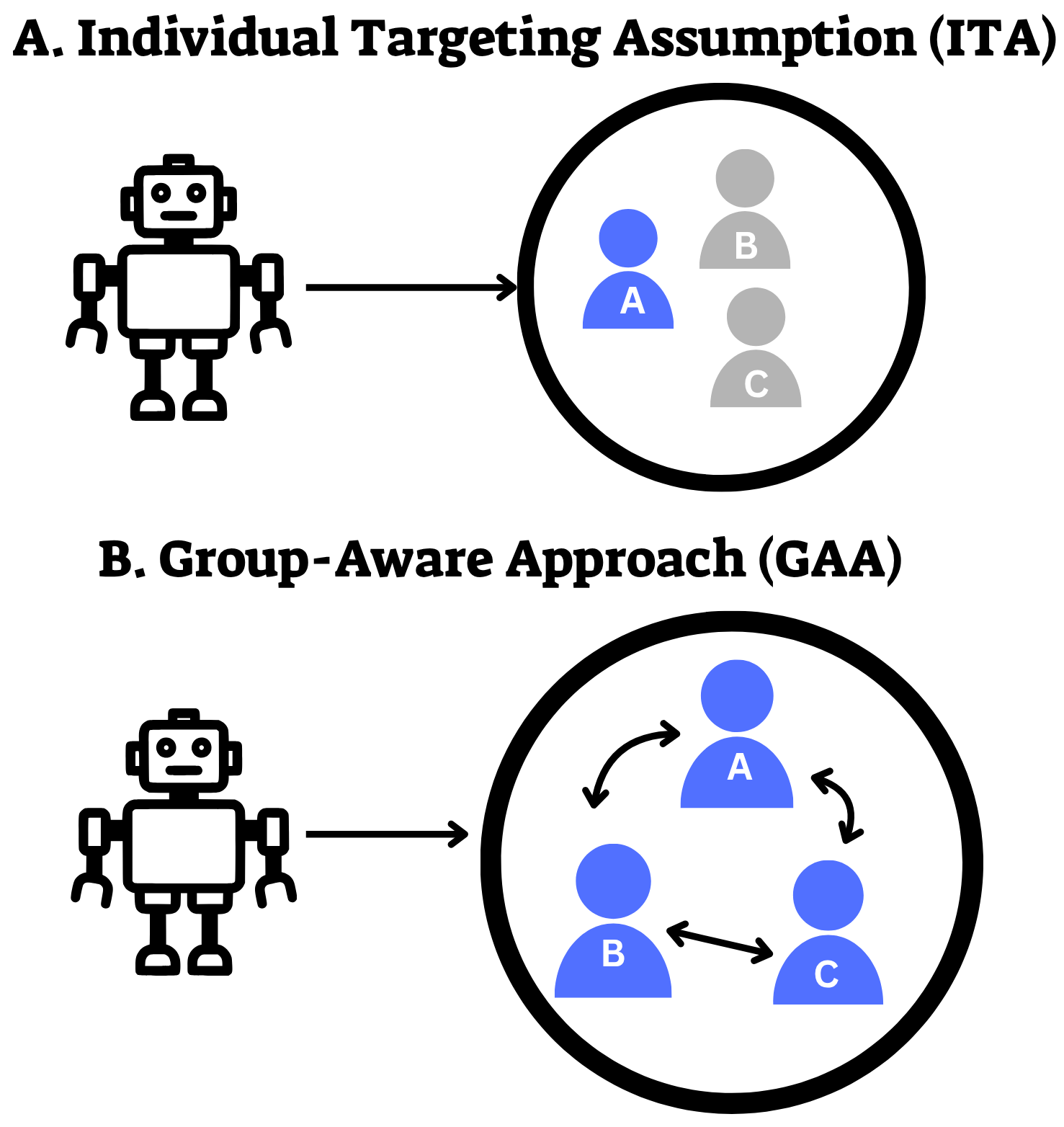}
    \caption{Two contrasting design paradigms in proactive HRI. (A) The Individual-targeting assumption (ITA): the robot models each co-present person as an independent engagement target. (B) The Group-aware approach (GAA): the robot treats co-present people as a cohesive social unit.}
    \label{fig:itv_group}
\end{figure}
\par On a spectrum, for the purpose of communication in HRI, robots can be broadly categorized from reactive to proactive~\cite{mok2016effects}.
Reactive robots respond to immediate user actions or environmental stimuli without planning or predicting future events~\cite{rozo2016learning}.
On the contrary, proactive robots initiate interactions and anticipate humans’ needs~\cite{van2024proactive, mok2016effects}.
Proactive HRI has emerged as a growing subfield within the past 15 years, driven by the desire for more human-robot interactions~\cite{van2024proactive}.
However, the vast majority of proactive HRI frameworks have been developed and evaluated in dyadic (one-to-one) settings, leaving the considerably more complex and common scenario of group interaction largely underexplored~\cite{van2024proactive}. 
This paper addresses that gap through a systematic literature review of proactive robot behaviour in human group settings.
\par The limited literature on proactive robots in human group settings reveals a core tension between individual-based interaction models used by robots and the group-based structure of real-world human social interaction~\cite{sebo2020robots, fischer2015initiating}. Proactive robots often focus on detecting and engaging individuals, despite Fischer et al.~\cite{fischer2015initiating} showing when a robot approached people sitting together, group members exchanged glances and negotiated attention before collectively turning toward the robot, a coordination step rarely accounted for. 
Two research streams partially address this challenge but remain largely disconnected. 
Crowd-aware proactive robot navigation has made progress in perceiving group movement patterns and formations to avoid collisions~\cite{mavrogiannis2023core}, but this spatial awareness rarely extends to understanding the social relationships, shared intentions, and joint attention that structure how groups actually behave. 
Moreover, HRI for social group interaction addresses turn-taking, addressee detection, and gaze allocation during interaction with humans~\cite{nakano2015generating, foster2017automatically, mutlu2009footing}, but predominantly in structured scenarios where the robot is already embedded in the group rather than proactively approaching from outside.
\par In this paper, we identify the assumption underlying these interaction challenges: that robots engage humans as individual targets rather than as members of a social group. 
We term this the \textit{Individual-Targeting Assumption (ITA)}, defined as the design tendency to model each person in a social group as an independent engagement target based solely on their individual signals (i.e., gaze, proximity, orientation, and speech) while remaining blind to the group's relational structure. 
Research work that explicitly models group-level properties to inform robot behaviour, we term the \textit{Group-Aware Approach (GAA)}.
Fig.~\ref{fig:itv_group} illustrates these two contrasting design paradigms.

In this systematic review, we examined whether proactive robots treat group members as individual targets or as a cohesive social unit. Drawing on inductive coding of 63 papers on proactive human-robot group interaction, the three research questions below were formulated from this coding process to organize the reported findings. We reviewed papers published between 2009 and 2025, given that from 2000 to 2008 no papers were found that address proactive human-robot group interaction. The three research questions are: (RQ1) How prevalent is the ITA and GAA across proactive human-robot group interactions? (RQ2) What failure modes arise when proactive robots apply individual targeting logic in human group settings? (RQ3) What group-aware techniques have been used in proactive HRI? 
Our analysis makes four primary contributions: (i) We identify and quantify the ITA as a structural assumption in proactive human-robot group interaction, present in approximately 60.3\% of the coded corpus across three forms: unreflective, critical, and transitional; (ii) we document three failure modes, \textit{engagement misdetection}, \textit{social ratification blindness}, and \textit{bystander neglect} (see fig. \ref{fig:failure_modes}); (iii) we catalogue the proactive robot’s system-level group-aware techniques in human-robot group interaction; and (iv) we identify \textit{group entry} as an interaction context in which robots detect, approach, and join social groups as entirely without \textit{group-aware behaviour} across the coded corpus, pointing to a critical underexplored design space.
\section{Related Work}
\subsection{Proactive HRI}
Proactive HRI refers to a broad range of robot capabilities where the robot takes charge in various situations, displays anticipatory behaviour, or offers assistance without waiting for an explicit command or prompt from the environment \cite{van2024proactive}. 
Proactive robots exercise these capabilities by perceiving individual-level social signals, such as gaze direction, proximity, and facial expression, as indicators of openness to interact with humans \cite{bohus2014directions, li2019inferring}. 
Bohus et al.'s \cite{bohus2014directions} directions robot and Kato et al.'s \cite{kato2015may} polite approaching robot exemplify this paradigm, both computing per-person intention scores from visual cues to decide when and how to approach them. 
More recently, researchers have improved proactive robots' capabilities through multimodal fusion \cite{li2019inferring} and state-based interaction management that profiles individual users to determine whether to approach or wait \cite{rajendran2023user}. 
However, scholars suggest that research on proactive HRI focuses more on dyadic interaction, leaving group interaction underexplored \cite{van2024proactive}.

\subsection{Robot in Human Group Interaction}
Robot-in-human group interaction involves two or more humans and at least one robot, where the social relationships among group members influence human-human and human-robot behaviour \cite{sebo2020robots}. 
Groups may vary  in whether they share common goals, are interdependent in their tasks, interact socially, or are only loosely associated \cite{sebo2020robots}. 
Kendon defines the F-formation as a spatial arrangement people adopt for face-to-face interaction, creating a shared space, the O-space, that both members and outsiders treat as socially bounded~\cite{kendon1990conducting, cristani2011social}.

Beyond spatial organization, group interaction is governed by participation roles. 
Goffman's distinguishes between \textit{ratified participants}, who are legitimately included in the interaction, and \textit{bystanders}, who are observers but not addressed as part of the main interaction \cite{goffman2008behavior, goffman1981forms}. 
These role distinctions inside a group carry direct implications for how a robot should allocate attention, gaze, and communicative initiative in a group~\cite{nakano2015generating, vazquez2017towards}. 
The roles are not fixed but dynamically negotiated. 
Fischer et al.~\cite{fischer2015initiating} demonstrated that when a robot approaches a group, members first exchange glances and negotiate attention with each other before any individual engages, a social ratification process entirely absent from dyadic interaction. 
This negotiation shows that group members constantly pay attention to each other’s behaviour. 
As a result, when a robot interacts with one person within a group, the interaction also affects other group members \cite{nielsen2023using}.
 
In group HRI, understanding human engagement signals becomes more complex than in dyadic settings because gaze and body orientation may reflect inter-human communication rather than robot-directed attention~\cite{zhang2021engagement, foster2017automatically, bohus2014directions}.
Furthermore, group HRI exhibits properties different from dyadic HRI. 
These include the formation of dominance hierarchies \cite{nakano2015generating}, participation imbalance~\cite{grassi2025strategies}, and the subgroup formation \cite{correia2018group}, which influence how new participants, human or robot, are perceived and accepted.  
Designing interaction for group HRI is therefore not an extension of dyadic HRI but a qualitatively different problem, which requires distinct perceptual models, social reasoning, and engagement strategies~\cite{schneiders2022non, abrams2020c, sebo2020robots}.

\section{Methodology}
We conducted a systematic literature review to investigate the prevalence and consequences of the ITA in proactive human-robot group settings, particularly in cases involving multiple human-robot interactions. Two researchers carried out the review and followed four stages consecutively: (i) database search, (ii) title and abstract screening, (iii) full-text review, and (iv) structured coding.
\subsection{Search Strategy}
\begin{figure}[!htbp]
    \centering
    \includegraphics[width=\columnwidth]{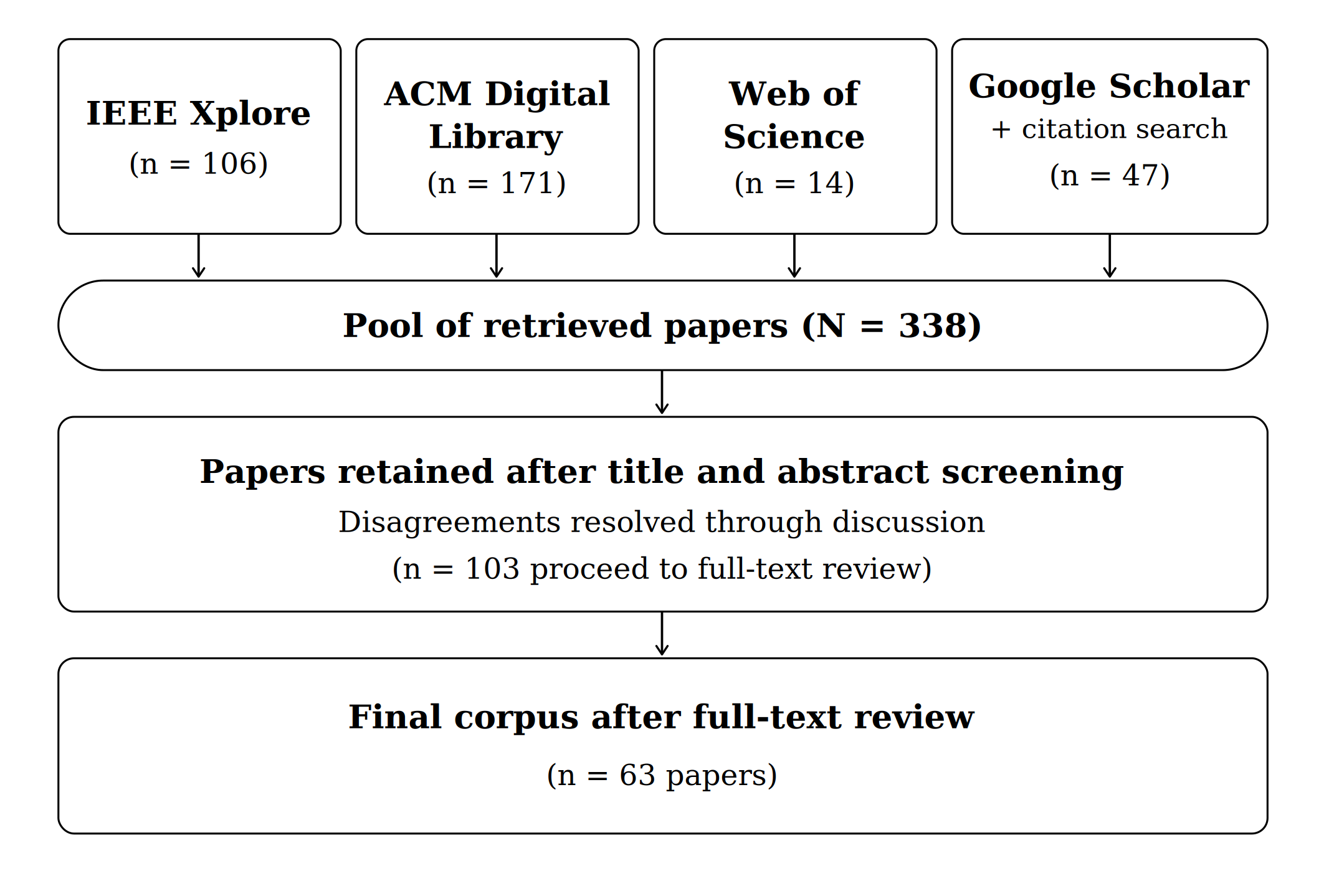}
    \caption{Screening and selection process for the systematic literature review.}
    \label{fig:screening}
\end{figure}

We searched three major academic databases: (i) IEEE Xplore, (ii) ACM Digital Library, and (iii) Web of Science (see Fig. \ref{fig:screening}).
These databases collectively cover the primary venues for publishing proactive HRI and social robotics research, including IEEE RO-MAN, HRI, IROS, ICSR, and ACM/CHI conferences and journals. 
Search queries combined terms related to three concepts: proactive robot behaviour (e.g., “proactive robot”, “robot initiation”, “robot approaching”), human group settings (e.g., “group”, “multiparty”, “team”, “social group”), and engagement (e.g., “engagement”, “interaction”, “approach”). 
Searches were performed from 2000 to 2025 to capture the full historical scope of the field. 
We also conducted Google Scholar searches, as well as searches of the publication lists of prominent authors in the proactive HRI field, and citations tracing back to research already identified in the corpus.

\subsection{Inclusion and exclusion Criteria}
We included papers that satisfied the following criteria: (i) The study involves at least one physical, simulated robot, or virtual agent (i.e., chatbot) interacting with a human group.
(ii) The robot demonstrates proactive behaviour, autonomously or through Wizard-of-Oz control, such as anticipating human or team needs, initiating or \mbox{managing} interaction, approaching, coordinating, and adjusting its \mbox{behaviour} in response to a multi-human situation.
(iii) Two or more humans are physically or virtually co-present in the environment during the interaction, and the robot perceives, reasons about, or acts within this multiparty context (e.g., selecting a person from a group, addressing multiple humans, managing attention among group members, or coordinating interaction with more than one human).
(iv) The navigation-based robots must be designed to interact with a member of the human group, not merely to navigate around humans to avoid obstacles.

We also considered the simulated robots and virtual agents that met the above-mentioned criteria, because proactive robots interacting with a human group are already scarce \cite{van2024proactive}. 
We excluded papers where robots exhibited only reactive behaviour; studies focused solely on software or architectural aspects without HRI concepts; studies limited to dyadic HRI; studies where robots interacted with multiple users only sequentially across separate sessions rather than with simultaneously co-present humans; multirobot systems operated by humans; and studies in which the robot only focused on environmental changes within a human group (e.g., adjusting room settings); and papers not written in English. 

\subsection{Screening Process}
We first combined all papers retrieved from the databases and supplementary searches into a single pool (N = 338). These originated from the ACM Digital Library (n = 171), Google Scholar (n = 47), IEEE Xplore (n = 106), and Web of Science (n = 14). We screened the papers in two stages. First, both researchers independently reviewed the titles and abstracts of all retrieved papers against the inclusion and exclusion criteria. When abstracts were insufficient to determine eligibility, the full text was briefly inspected. Each paper received an independent include or exclude decision from each researcher. To assess the reliability of this stage, we calculated inter-rater agreement using Cohen's Kappa $\kappa$ \cite{cohen1960coefficient}. 
The reviewers achieved $\kappa = 0.812$, indicating almost perfect agreement \cite{landis1977measurement}. The reviewers resolved the disagreements through discussion. After resolving disagreements, 103 papers remained. 
Second, these papers underwent full-text review, resulting in a final corpus of 63 papers included in the structured coding and analysis. 
The reviewer found the earliest paper meeting the inclusion criteria dated to 2009.
\subsection{Analytic Approach and Coding}
Two authors adopted conventional content analysis, in which coding categories emerged inductively through~\mbox{repeated} engagement with the data rather than being predetermined based on existing theory or prior research~\cite{hsieh2005three}.
We started the review with a general interest in how proactive robots handle group contexts and did not specify the ITA construct, its failure modes, or the interaction context classification in advance.
We finalized our research questions after the inductive coding process based on the five themes that we generated in two steps. First, we annotated and subcoded recurring patterns and robot behaviours independently and interpretively within each paper. In the second step, based on those subcodes, we iteratively compared those subcodes across the corpus, and after rounds of comparison and refinement by both authors, semantically related subcodes were consolidated into the five overarching themes. Disagreements at both the subcoding and theme levels were resolved through discussion until consensus was reached, which is consistent with the iterative nature of conventional content analysis. The full corpus and codebook are provided as supplementary materials.\footnote{OSF repository link: \url{https://osf.io/kgqht/}.}
\section{Findings}
Our final coded corpus comprised $N = 63$ papers spanning publication years 2000 to 2025. 
Venues represented include HRI, RO-MAN, CHI, THRI, IJSR, ICSR, AAMAS, RA-L, and CSCW. 
During the full-text review, we annotated patterns of interaction to identify how proactive robots modelled, or failed to model, the social structure of human groups. Our coding scheme comprised five themes, the first of which emerged as a recurring pattern we termed the Individual-targeting assumption (ITA). We then formalised the remaining categories into four additional themes using the same inductive process. 

Settings of the evaluated ranged from laboratory experiments to real-world field deployments, with the remainder using a conceptual or simulation format. Group configurations ranged from two humans to unspecified crowds; the majority of papers described a 1-robot-to-$N$-humans topology without specifying $N$.
Among studies specifying a fixed group size, the most common configurations were pairs (1R:2H, $n = 7$) and triads (1R:3H, $n = 6$). Tables~\ref{tab:ita_class}--\ref{tab:techniques} summarise the four coded themes. 
RQ(i) is addressed by Tables~\ref{tab:ita_class} and~\ref{tab:phases}, RQ(ii) by Table~\ref{tab:failure_modes}, and RQ(iii) by Table~\ref{tab:techniques}. 
We have discussed the nature of the themes below, along with the tables.
\begin{table}[!tb]
\centering
\caption{ ITA Classification \& GAA ($N = 63$) Code Distribution}
\label{tab:ita_class}
\setlength{\tabcolsep}{2.5pt}
\renewcommand{\arraystretch}{1.25}
\scriptsize
\begin{tabularx}{\columnwidth}{L{1.4cm} C{0.75cm} X L{1.6cm}}
\toprule
\rowcolor{hdrblue}
\textbf{ITA \& GAA Codes} & \textbf{$n$ (\%)} & \textbf{Definition} & \textbf{Examples} \\
\midrule
Unreflective ITA
  & 20 (31.7)
  & \textit{Individual-targeting assumption} applied; no group-level limitation acknowledged.
  & \cite{hoque2014empirical}; \cite{gyoda2011mobile}; \cite{chen2023outperformance} \\
\rowcolor{rowgray}
Critical ITA
  & 15 (23.8)
  & \textit{Individual-targeting assumption} applied; group-level gap explicitly flagged in
    discussion or future work.
  & \cite{kobayashi2011assisted}; \cite{bohus2014directions}; \cite{haripriyan2024human} \\
Transitional
  & 3 (4.8)
  & ITA breakdown documented; no GAA implemented.
  & \cite{fischer2015initiating}; 
  \cite{taylor2025rapidly}; 
  \cite{pusceddu2025will} \\
\rowcolor{rowgray}
Group-Aware Behaviour
  & 29 (46.0)
  & Models ${\geq}1$ relational group property (F-formation, speaking
    time, inter-human gaze) to inform proactive behaviour.
  & \cite{tennent2019micbot}; \cite{grassi2025strategies}; \cite{gillet2022learning} \\
\midrule
\textbf{ITA Total}
  & \textbf{38 (60.3)}
  & \multicolumn{2}{l}{Unreflective + Critical + Transitional} \\
\bottomrule
\multicolumn{4}{p{\dimexpr\columnwidth-6pt}}{%
  \tiny\textit{Note:} Four papers~(\cite{love2024towards};
  \cite{bohus2009models};
  \cite{bohus2010facilitating};
  \cite{mutlu2009footing}) are dual-coded
  as both ITA and group-aware behaviour.} \\
\end{tabularx}
\end{table}

\textbf{ITA Classification (theme 1)} captures a paper's relationship to individual targeting along a spectrum of awareness: papers in which the robot uses individual targeting logic in human-robot group interaction and the authors do not acknowledge any group-level limitation \textit{(Unreflective ITA)}; papers in which the same logic is used but the authors explicitly flag the group-level gap in their discussion or future-work sections \textit{(Critical ITA)}; and papers that empirically document the breakdown of individual targeting and articulate a group-level need without implementing a full group-aware solution \textit{(Transitional)}.
The \textit{Transitional} code was not anticipated and emerged mid-review from papers that documented the breakdown of individual targeting without implementing a group-aware solution, consistent with conventional content analysis, which allows new categories to surface from the data \cite{hsieh2005three}. 
%
%
\begin{table}[!tb]
\centering
\caption{ITA \& GAA Distribution Across Interaction Context}
\label{tab:phases}
\setlength{\tabcolsep}{2.5pt}
\renewcommand{\arraystretch}{1.25}
\scriptsize
\begin{tabularx}{\columnwidth}{L{1.3cm} C{0.72cm} C{0.72cm} C{0.55cm} X}
\toprule
\rowcolor{hdrblue}
\textbf{Interaction Context Codes} & \textbf{$n$ (\%)} & \textbf{GAA $n$} & \textbf{ITA $n$} & \textbf{Key Observation} \\
\midrule
Facilitation
  & 46 (73.0) & 28 & 21
  & Only context with substantial GA solutions; robot is already embedded in
    ongoing interaction. \\
\rowcolor{rowgray}
Group Entry
  & 24 (38.1) & 4 & 22
  & Four GAA papers involve \textit{group entry} but apply their group-aware techniques to \textit{turn-taking} and \textit{dialogue management} once interaction has begun, not to how the robot reads or enters the group. Only Joosse et al. \cite{joosse2014cultural} apply \textit{group-aware behaviour} (F-formation; relational geometry) to the spatial conduct. No paper implements \textit{group-aware behaviour} in \textit{group entry} context within the proactive HRI literature. \\
Spatial Conduct
  & 8 (12.7) & 2 & 6
  & GAA papers involve spatial conduct within multi-phase interactions;
    no purely spatial GAA solution exists. \\
\bottomrule
\multicolumn{5}{p{\dimexpr\columnwidth-6pt}}{%
  \tiny\textit{Note:} 14 papers span multiple contexts; row totals, therefore, exceed $N = 63$. Percentages are of $N = 63$.} \\
\end{tabularx}
\end{table}

\textbf{The Group-Aware Approach (GAA) (theme 2)} comprises papers coded as \textit{group-aware behaviour}. 
The theme addresses papers that explicitly model at least one relational property of the group as a social unit to inform proactive behaviour, departing from individual targeting logic rather than merely acknowledging its limits. 
Four papers were dual-coded, in which ITA governed one interaction context and GAA logic governed the other.

\textbf{The Interaction Context (theme 3)} describes the stage of the human–robot group encounter in which the proactive behaviour occurs.
Across the corpus, three distinct interaction contexts emerged. 
\textit{Group entry} describes encounters in which the robot is outside an existing social group and proactively acts to gain entry either by detecting and initiating contact with potential interlocutors or by navigating the internal social boundaries of a pre-formed group.
\textit{Facilitation} describes encounters in which the robot is already embedded in an ongoing multi-party interaction and proactively manages its dynamics (turn-taking, participation balance, conflict resolution, and gaze allocation).
\textit{Spatial conduct} covers physical movement and positioning in shared space where the presence or structure of human groups shapes the robot's spatial behaviour, whether or not engagement is the primary goal.
\begin{figure}[!htbp]
    \centering
    \includegraphics[width=\columnwidth]{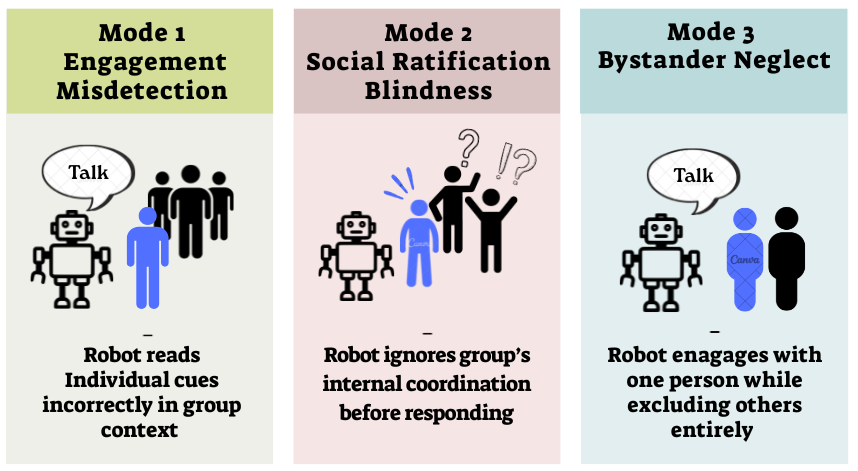}
    \caption{Three failure modes arising from the Individual-targeting assumption (ITA) in proactive human-robot group interaction.} 
    \label{fig:failure_modes}
\end{figure}

We coded \textbf{Failure Mode (theme 4)} when a paper reported breakdowns in group settings that could be interpreted as related to individual targeting. 
Three failure types emerged inductively (see fig. \ref{fig:failure_modes}). 
\textit{Engagement misdetection} refers to cases in which the robot misinterprets a signal exchanged between group members (e.g., a head turn toward a companion) as evidence of disengagement. 
\textit{Social ratification blindness} describes cases in which the robot addresses an individual before the group has had the opportunity to negotiate internally whether to engage, a collective decision that cannot be inferred from any individual's signals alone.
\textit{Bystander neglect} describes cases where a non-targeted group member experiences discomfort, social exclusion, or disrupted participation. 
Since most papers do not foreground group-level consequences of their design, this theme required interpretive judgment from both coders \cite{krippendorff2018content}.
%
%
\begin{table}[!tb]
\centering
\caption{Failure Modes observed  in Proactive Group HRI ($N = 63$)}
\label{tab:failure_modes}
\setlength{\tabcolsep}{3pt}
\renewcommand{\arraystretch}{1.25}
\scriptsize
\begin{tabularx}{\columnwidth}{L{1.7cm} X C{0.35cm}}
\toprule
\rowcolor{hdrblue}
\textbf{Failure Modes} & \textbf{Definition, Consequence \& Key Evidence} & \textbf{$n$} \\
\midrule
Engagement Misdetection
  & Robot misinterprets intergroup cues
(e.g., lateral gaze toward a companion) as individual disengagement.
Errors nearly doubled in multiparty vs. dyadic
settings~\cite{bohus2014directions}; gaze toward a companion is misread as disengagement~\cite{zhang2021engagement}; individual cues
produce classifier jitter without relational
context~\cite{foster2017automatically}.
  & 6 \\
\rowcolor{rowgray}
Social Ratification Blindness
  & Robot bypasses the group's internal agreement process and addresses one member directly, forcing the group to self-organise around the robot's limitations. Members exchanged glances before orienting to the robot ~\cite{fischer2015initiating}; teams created a self-appointed robot manager ~\cite{haripriyan2024human}; robot authority was overridden by group dynamics~\cite{pusceddu2025will}.
  & 9 \\
Bystander Neglect
  & Robot exclusively engages one member from the group;
others experience exclusion, reduced participation, or
disrupted dynamics. Targeted users broke the robot's interaction
structure to re-include their ignored companion~\cite{mutlu2009footing};
individual-model repair disrupted group interaction and yielded worse
recall than a non-proactive condition~\cite{leite2016autonomous};
an uncoordinated approach caused severe obstruction among non-targeted team members~\cite{taylor2025rapidly}.
  & 18 \\
\bottomrule
\multicolumn{3}{@{}p{\columnwidth}@{}}{%
  \tiny\textit{Note:} Failure modes span both ITA and
  GAA papers; several GAA papers document failure modes as a motivated baseline before presenting their solution;  in these cases, both codes were applied. Papers may exhibit
  multiple failure modes.} \\
\end{tabularx}
\end{table}
%
\begin{table}[!tb]
\centering
\caption{Group-Aware Techniques Across 29 GAA Papers}
\label{tab:techniques}
\setlength{\tabcolsep}{2.5pt}
\renewcommand{\arraystretch}{1.11}
\scriptsize
\begin{tabularx}{\columnwidth}{L{1.55cm} X C{0.4cm}}
\toprule
\rowcolor{hdrblue}
\textbf{Technique} & \textbf{Definition \& Sample Papers} & \textbf{$n$} \\
\midrule
\multicolumn{3}{l}{\cellcolor{rowgray}\textit{Participation Management}} \\
Group turn-taking management
  & Manages floor allocation and turn transitions across all group members collectively. \cite{pham2024embodied}; \cite{shamekhi2019multimodal}; \cite{tennent2019micbot}
  & 18 \\
Participation balance
  & Tracks relative participation across members and intervenes to restore equity. \cite{pham2024embodied}; \cite{shamekhi2019multimodal}
  & 14 \\
Speaking-time tracking
  & Monitors the cumulative floor time for each participant and uses imbalance as a trigger.
    \cite{shamekhi2019multimodal}; \cite{tennent2019micbot}
  & 10 \\
Dominance ranking
  & Models conversational dominance to counteract power imbalances. \cite{grassi2025strategies}
  & 6 \\
\midrule
\multicolumn{3}{l}{\cellcolor{rowgray}\textit{Group Structure Sensing}} \\
Group-level state
  & Tracks a holistic measure of the group's current state (e.g., engagement, affect, activity) as a trigger for proactive behaviour.
    \cite{jamshad2024taking}
  & 9 \\
Inter-human gaze~monitoring
  & Monitors gaze exchanges between human participants, not just human-to-robot gaze, to understand group dynamics. 
    \cite{love2024towards}; \cite{correia2023robotic}; \cite{bohus2010facilitating}
  & 3 \\
Relational geometry
  & Uses geometric features of the group as a whole (centroid, mean facing direction) to plan approach or gaze. \cite{vazquez2017towards}; 
  & 3 \\
F-formation awareness
  & Detects or respects Kendon F-formations to understand group spatial structure.\cite{vazquez2017towards}
  & 2 \\
Subgroup detection
  & Identifies conversational subgroups within the larger group (e.g., via Louvain algorithm on interaction graphs).
    \cite{grassi2025strategies}
  & 1 \\
\midrule
\multicolumn{3}{l}{\cellcolor{rowgray}\textit{Social-Affective}} \\
Group backchanneling
  & Produces backchannels (nods, brief affirmations) directed at multiple group members simultaneously.
    \cite{pham2024embodied}; \cite{tennent2019micbot}; \cite{ochi2025robot}
  & 5 \\
Group-level emotion
  & Group-level emotion or in-group framing build solidarity.
    \cite{correia2018group}; \cite{jung2015using}; 
  & 3 \\
Relationship repair
  & Proactively detects and mediates interpersonal conflict or norm violations within a group.
    \cite{jung2015using}; \cite{chen2025maintaining}
  & 2 \\
\midrule
\multicolumn{3}{l}{\cellcolor{rowgray}\textit{Trust Modelling}} \\
Group trust modeling
  & Models trust at the team or group level rather than only as a dyadic property.
    \cite{correia2018group}
  & 1 \\
\midrule
\multicolumn{3}{@{}p{\columnwidth}@{}}{%
  \tiny\textit{Note:} Of the four GAA papers spanning \textit{group entry}, three apply individual targeting logic at the moment of entry and one~\cite{joosse2014cultural} addresses \textit{spatial conduct} only via F-formation awareness and relational geometry. No group-aware technique addresses how a robot reads whether a pre-formed group is open to being approached before it initiates contact. $^{*}$One paper may apply multiple techniques; total applications: 77.} \\
\end{tabularx}
\end{table}

\textbf{Group-Aware Technique (Theme 5)} emerged inductively from papers coded as \textit{group-aware behaviour} (related to Theme 2) and captures the specific group-level mechanism a system uses to detect, model, or regulate interaction dynamics.
Codes include speaking-time tracking, F-formation awareness, subgroup detection, and inter-human gaze monitoring.

%

\section{Discussion}
The review of 63 papers addressed three research questions that organized the analysis: the prevalence of the ITA and GAA (RQ1), the failure modes produced by ITA in group settings (RQ2), and the group-aware techniques applied in proactive group HRI (RQ3). 
The findings suggest that ITA is the dominant design paradigm in group HRI (60.3\% of the coded corpus) (RQ1). 
Three failure modes (\textit{engagement misdetection, social ratification blindness,} and \textit{bystander neglect}) are recurring and empirically observed as outcomes of ITA, but in a few cases can also be found in GAA papers, sometimes as a motivating baseline for the proposed solution (RQ2).  Group-aware techniques (RQ3) appear almost exclusively in the \textit{facilitation} context. Among the four GAA papers spanning \textit{group entry}, three rely on individual targeting logic at the point of robot entry, and one focuses only on \textit{spatial conduct} by the robot, leaving how a robot should detect, interpret, and negotiate entry into a pre-formed group as a collective social unit unexplored. Taken together, these findings reframe proactive human-robot group interaction, particularly in the \textit{group entry} context, not as an extension of dyadic HRI but as a qualitatively distinctive design problem.

\subsection{ITA Design Tendency: Dyadic Interaction Within Group Settings}
We observed that ITA is a design tendency in which a robot targets a single person in a group setting, effectively reproducing a dyadic interaction paradigm within a multi-human context. This result is consistent with and extends broader evidence that dyadic paradigms dominate HRI research. For example, even in recent studies, Schneiders et al. \cite{schneiders2022non} found that only 28\% of the 587 HRI papers investigated non-dyadic interaction. Sebo et al. \cite[p.~176.1]{sebo2020robots} similarly observed that HRI research has “largely focused on dyadic interactions” and argued the field must consider a robot's influence at the group level. This pattern extends to proactive HRI, where group interaction remains largely underexplored \cite{van2024proactive}; yet no prior work has quantified how deeply individual targeting logic is embedded in this literature. To the best of our knowledge, our findings provide the first quantitative estimate of ITA prevalence specifically within the proactive HRI literature in group settings. 
 
\subsection{The \textit{Group Entry} Gap: The Field's Most Pressing Open Challenge}
A salient finding is ITA’s concentration in the \textit{group entry} context. 22 of 25 papers coded as \textit{group entry} demonstrate that robots rely on individual targeting, while in the few, robots demonstrate \textit{group-aware behaviour} only after joining the group. This gap reflects the absence of mechanisms for socially negotiating entry, even though theory on F-formations and participation ratification shows that group membership requires collective sanction. As real-world robots routinely encounter pre-formed groups, the field has largely addressed the \textit{facilitation} inside groups, but not how robots should join them while avoiding ITA. 
This is because, during \textit{facilitation}, robots observe group structure progressively over an extended interaction, whereas \textit{group entry} requires inferring F-formations, inter-human gaze, group boundaries, and group openness cues from a brief observation window before any interaction has begun. This makes \textit{group entry} a harder and more time-constrained robot perception problem than the post-entry case.

\subsection{Failure Modes as a Causal Chain}
The three failure modes documented in our corpus, \textit{engagement misdetection} (n = 6), \textit{social ratification blindness} (n = 9), and \textit{bystander neglect} (n = 18), are not independent problems. They form a causal chain rooted in a single structural error by indicating the absence of a relational model of the group. Misdetection of engagement signals leads a robot to misread group readiness, bypassing the group's internal ratification process, which in turn results in the exclusion of non-targeted members.

\textit{Engagement misdetection} arises because individual cues such as gaze, body orientation, and proximity are inherently ambiguous for robots in group HRI contexts. Vázquez et al. \cite{vazquez2017towards} show that gaze and body orientation signals in group conversations must be interpreted jointly rather than independently. Similarly, Oertel et al. \cite{oertel2020engagement} note that prior engagement models largely assume dyadic interaction in human-agent interaction. However, multiparty interaction requires additional mechanisms to capture group dynamics, such as engagement density and participation harmony,  which cannot be inferred from individual signals alone \cite{matsuyama2015four}. 
Empirically, Bohus et al. \cite{bohus2014directions} report that engagement error rates roughly doubled in multiparty settings compared to dyadic ones. 
For example, lateral gaze reverses in meaning, as what signals disengagement in a dyadic setting can indicate continued engagement through another participant in a group. 
Without modelling inter-human relations, robots cannot disambiguate these cues. 

\textit{Social ratification blindness} occurs when a robot addresses an individual before the group has negotiated internally whether to engage at all. 
Pillet-Shore’s conversation analysis documents that, in natural interaction, entering a group is a negotiated achievement in which newcomers are typically included through offers of participation rather than requests to participate \cite{pillet2010making}. So, robots should not assertively interrupt groups but proactively detect openness, signal availability, and wait for or elicit group-level ratification before addressing an individual. Robots that bypass this negotiation violate the social order of entry even when individual engagement detection is correct. Studies have shown that group members negotiated attention among themselves before engaging a robotic trash can \cite{fischer2015initiating}. Research in HRI further demonstrates that bypassing ratified participants while engaging a single member reduces perceived group cohesion among non-addressed members \cite{mutlu2009footing}. 

\textit{Bystander neglect}, the most common failure mode, occurs when a robot exclusively engages with one group member, thereby resulting in exclusion and unequal participation. Because non-targeted members generate no robot-directed signals, individual-targeting models cannot detect this form of exclusion. A robot's social behaviour shapes human–human conversational dynamics within a team, with inclusive or supportive robot behaviours producing more equitable participation among members \cite{traeger2020vulnerable, tennent2019micbot, gillet2021robot}, while selective or unequal robot engagement can lead to participation imbalances and undermine inclusion \cite{sebo2020robots}. Mongile et al. \cite {mongile2023if} further show that when a robot selectively excludes a group member, included participants attempt to compensate by redirecting their choices toward the excluded peer.

\subsection{Group-Aware Techniques: Promise and Barriers}
In our corpus, 13 group-aware techniques across 77 applications in 29 papers show substantial progress in the \textit{facilitation} context.  These 13 group-aware techniques cluster into four functional categories:  \textit{participation management, group structure sensing, social-affective regulation}, and \textit{trust modelling} (see \ref{tab:techniques}). Techniques such as group turn-taking management (n = 19), participation balance tracking (n = 14), and speaking-time monitoring (n = 10) address participation inequities created by individual targeting in embedded group interactions. Several studies demonstrate the effectiveness of these approaches. For instance, Gillet et al. \cite{gillet2022learning} show that imitation learning can produce gaze policies that encourage more active participation in group HRI without compromising interaction quality.  Shamekhi and Bickmore \cite{shamekhi2019multimodal} develop a multimodal facilitation framework for managing group decision-making. Similarly, Matsuyama et al. \cite{matsuyama2015four} treat group engagement density as a system variable requiring active management. Collectively, these contributions represent genuine advances in group-aware techniques in the \textit{facilitation} context.

These group-aware techniques do not transfer to the \textit{group entry} context because the information available to the robot differs fundamentally, not because the problem has been ignored. During \textit{facilitation}, robots observe participation patterns, spatial configurations, and interaction dynamics over time, making group structure progressively measurable. During \textit{group entry}, robots must infer the same structure from brief and noisy observations before any interaction has begun. Hedayati and Szafir \cite{hedayati2022predicting} show that even robots already situated within a group face systematic errors in estimating group size and participant positions due to occlusions and sensor noise. While systems such as REFORM improve F-formation detection over heuristic baselines \cite{hedayati2020reform}, detection errors persist, and robot behaviours such as approach positioning and engagement initiation that depend on accurate group perception remain at risk.

\section{Limitations and Future Scope} 
Several limitations constrain the interpretation of this review.
Although inter-rater agreement for study inclusion was high ($\kappa = 0.812$), we did not compute a formal reliability metric for finding the overarching themes but relied on iterative, consensus-based decision.
The corpus spans 25 years, but we did not analyze temporal trends, preventing conclusions about whether ITA is declining or GAA methods are increasing. Additionally, the threshold for defining \textit{group-aware behaviour} was intentionally inclusive, grouping approaches of varying depth. Finally, most studies in our corpus occurred in laboratories, limiting insight into complex real-world group dynamics.

Despite these constraints, the review contributes several key insights to HRI. It reveals a structural pattern of proactive group HRI in which robots overwhelmingly rely on individual targeting logic, particularly during the \textit{group entry} context. The synthesis also identifies three recurring failure modes that collectively point to a deeper design limitation, namely, the absence of relational models that represent the group as a social unit. At the same time, the catalogue of group-aware techniques shows that meaningful progress has been made in the \textit{facilitation} context, where robots can manage participation and turn-taking once embedded in an interaction.

The review addresses a critical gap by demonstrating that the unresolved problem is not group interaction in general, but how robots detect, interpret, and negotiate entry into a group as a collective social unit. Future research should therefore prioritise group-entry perception, including robot-centric detection of F-formations, inter-member gaze, and signals of group openness during approach. It should also develop group-level reasoning and evaluation frameworks that test ratification, inclusion, and bystander effects, so that proactive robots can move from dyadic competence toward socially legitimate participation in real-world group settings.

\bibliographystyle{IEEEtran}
\bibliography{reference_old}

@article{sebo2020robots,
  title={Robots in groups and teams: a literature review},
  author={Sebo, Sarah and Stoll, Brett and Scassellati, Brian and Jung, Malte F},
  journal={Proceedings of the ACM on Human-Computer Interaction},
  volume={4},
  number={CSCW2},
  pages={1--36},
  year={2020},
  publisher={ACM New York, NY, USA}
}

@inproceedings{bohus2014directions,
  title={Directions robot: in-the-wild experiences and lessons learned},
  author={Bohus, Dan and Saw, Chit W and Horvitz, Eric},
  booktitle={Proceedings of the 2014 international conference on Autonomous agents and multi-agent systems},
  pages={637--644},
  year={2014}
}

@inproceedings{fischer2015initiating,
  title={Initiating Interactions and Negotiating Approach: A Robotic Trash Can in the Field.},
  author={Fischer, Kerstin and Yang, Stephen and others},
  booktitle={AAAI spring symposia},
  year={2015}
}

@article{mavrogiannis2023core,
  title={Core challenges of social robot navigation: A survey},
  author={Mavrogiannis, Christoforos and Baldini, Francesca and others},
  journal={ACM Transactions on Human-Robot Interaction},
  volume={12},
  number={3},
  pages={1--39},
  year={2023},
  publisher={ACM New York, NY}
}

@article{nielsen2023using,
  title={Using user-generated youtube videos to understand unguided interactions with robots in public places},
  author={Nielsen, Sara and Skov, Mikael B and Hansen, Karl Damkj{\ae}r and Kaszowska, Aleksandra},
  journal={ACM Transactions on Human-Robot Interaction},
  volume={12},
  number={1},
  pages={1--40},
  year={2023},
  publisher={ACM New York, NY}
}

@article{van2024proactive,
  title={What is proactive human-robot interaction?-a review of a progressive field and its definitions},
  author={van Den Broek, Marike Koch and Moeslund, Thomas B},
  journal={ACM Transactions on Human-Robot Interaction},
  volume={13},
  number={4},
  pages={1--30},
  year={2024},
  publisher={ACM New York, NY}
}

@inproceedings{gomez2025developing,
  title={Developing Robots for Society},
  author={Gomez, Randy},
  booktitle={2025 20th ACM/IEEE International Conference on Human-Robot Interaction (HRI)},
  pages={2--2},
  year={2025},
  organization={IEEE}
}

@article{li2019inferring,
  title={Inferring user intent to interact with a public service robot using bimodal information analysis},
  author={Li, Kang and Sun, Shiying and others},
  journal={Advanced Robotics},
  volume={33},
  number={7-8},
  pages={369--387},
  year={2019},
  publisher={Taylor \& Francis}
}

@inproceedings{kato2015may,
  title={May I help you? Design of human-like polite approaching behavior},
  author={Kato, Yusuke and Kanda, Takayuki and Ishiguro, Hiroshi},
  booktitle={Proceedings of the Tenth Annual ACM/IEEE International Conference on Human-Robot Interaction},
  pages={35--42},
  year={2015}
}

@inproceedings{rajendran2023user,
  title={User profiling based proactive interaction manager for adaptive human-robot interaction},
  author={Rajendran, Hoashalarajh and Bandara, HM Ravindu T and Jayasekara, AGBP and Chandima, DP},
  booktitle={2023 Moratuwa Engineering Research Conference (MERCon)},
  pages={632--637},
  year={2023},
  organization={IEEE}
}

@book{kendon1990conducting,
  title={Conducting interaction: Patterns of behavior in focused encounters},
  author={Kendon, Adam},
  volume={7},
  year={1990},
  publisher={CUP Archive}
}

@inproceedings{cristani2011social,
  title={Social interaction discovery by statistical analysis of f-formations.},
  author={Cristani, Marco and Bazzani, Loris and others},
  booktitle={BMVC},
  volume={2},
  number={4},
  pages={10--5244},
  year={2011}
}

@book{goffman2008behavior,
  title={Behavior in public places},
  author={Goffman, Erving},
  year={2008},
  publisher={Simon and Schuster}
}

@book{goffman1981forms,
  title={Forms of talk},
  author={Goffman, Erving},
  year={1981},
  publisher={University of Pennsylvania Press}
}

@article{nakano2015generating,
  title={Generating robot gaze on the basis of participation roles and dominance estimation in multiparty interaction},
  author={Nakano, Yukiko I and Yoshino, Takashi and Yatsushiro, Misato and Takase, Yutaka},
  journal={ACM Transactions on Interactive Intelligent Systems (TiiS)},
  volume={5},
  number={4},
  pages={1--23},
  year={2015},
  publisher={ACM New York, NY, USA}
}

@inproceedings{jamshad2024taking,
  title={Taking initiative in human-robot action teams: How proactive robot behaviors affect teamwork},
  author={Jamshad, Rabeya and Haripriyan, Arthi and others},
  booktitle={Companion of the 2024 ACM/IEEE International Conference on Human-Robot Interaction},
  pages={559--562},
  year={2024}
}

@inproceedings{vazquez2017towards,
  title={Towards robot autonomy in group conversations: Understanding the effects of body orientation and gaze},
  author={V{\'a}zquez, Marynel and Carter, Elizabeth J and McDorman, Braden and Forlizzi, Jodi and Steinfeld, Aaron and Hudson, Scott E},
  booktitle={Proceedings of the 2017 ACM/IEEE International Conference on Human-Robot Interaction},
  pages={42--52},
  year={2017}
}

@inproceedings{zhang2021engagement,
  title={Engagement intention estimation in multiparty human-robot interaction},
  author={Zhang, Zhijie and Zheng, Jianmin and Thalmann, Nadia Magnenat},
  booktitle={2021 30th IEEE international conference on robot \& human interactive communication (RO-MAN)},
  pages={117--122},
  year={2021},
  organization={IEEE}
}

@article{foster2017automatically,
  title={Automatically classifying user engagement for dynamic multi-party human--robot interaction},
  author={Foster, Mary Ellen and Gaschler, Andre and Giuliani, Manuel},
  journal={International Journal of Social Robotics},
  volume={9},
  number={5},
  pages={659--674},
  year={2017},
  publisher={Springer}
}

@article{grassi2025strategies,
  title={Strategies for Controlling the Conversation Dynamics in Multi-Party Human-Robot Interaction},
  author={Grassi, Lucrezia and Recchiuto, Carmine Tommaso and Sgorbissa, Antonio},
  journal={International Journal of Social Robotics},
  volume={17},
  number={8},
  pages={1517--1539},
  year={2025},
  publisher={Springer}
}

@inproceedings{correia2018group,
  title={Group-based emotions in teams of humans and robots},
  author={Correia, Filipa and Mascarenhas, Samuel and others},
  booktitle={Proceedings of the 2018 ACM/IEEE international conference on human-robot interaction},
  pages={261--269},
  year={2018}
}

@inproceedings{schneiders2022non,
  title={Non-dyadic human-robot interaction: Concepts and interaction techniques},
  author={Schneiders, Eike},
  booktitle={2022 17th ACM/IEEE International Conference on Human-Robot Interaction (HRI)},
  pages={1176--1178},
  year={2022},
  organization={IEEE}
}

@article{abrams2020c,
  title={I--C--E framework: Concepts for group dynamics research in human-robot interaction: Revisiting theory from social psychology on ingroup identification (I), cohesion (C) and entitativity (E)},
  author={Abrams, Anna MH and der P{\"u}tten, Astrid M Rosenthal-von},
  journal={International Journal of Social Robotics},
  volume={12},
  number={6},
  pages={1213--1229},
  year={2020},
  publisher={Springer}
}

@article{cohen1960coefficient,
  title={A coefficient of agreement for nominal scales},
  author={Cohen, Jacob},
  journal={Educational and psychological measurement},
  volume={20},
  number={1},
  pages={37--46},
  year={1960},
  publisher={Sage Publications Sage CA: Thousand Oaks, CA}
}

@article{landis1977measurement,
  title={The measurement of observer agreement for categorical data},
  author={Landis, J Richard and Koch, Gary G},
  journal={biometrics},
  pages={159--174},
  year={1977},
  publisher={JSTOR}
}

@article{hsieh2005three,
  title={Three approaches to qualitative content analysis},
  author={Hsieh, Hsiu-Fang and Shannon, Sarah E},
  journal={Qualitative health research},
  volume={15},
  number={9},
  pages={1277--1288},
  year={2005},
  publisher={Sage Publications Sage CA: Thousand Oaks, CA}
}

@book{krippendorff2018content,
  title={Content analysis: An introduction to its methodology},
  author={Krippendorff, Klaus},
  year={2018},
  publisher={Sage publications}
}

@inproceedings{bohus2009models,
  title={Models for multiparty engagement in open-world dialog},
  author={Bohus, Dan and Horvitz, Eric},
  booktitle={Proceedings of the SIGDIAL 2009 Conference},
  pages={225--234},
  year={2009}
}

@inproceedings{haripriyan2024human,
  title={Human-robot action teams: A behavioral analysis of team dynamics},
  author={Haripriyan, Arthi and Jamshad, Rabeya and Ramaraj, Preeti and Riek, Laurel D},
  booktitle={2024 33rd IEEE International Conference on Robot and Human Interactive Communication (ROMAN)},
  pages={1443--1448},
  year={2024},
  organization={IEEE}
}

@article{pusceddu2025will,
  title={Who will you imitate? Studying reciprocal influence in children-robot groups during an imitation game},
  author={Pusceddu, Giulia and Cocchella, Francesca and others},
  journal={Frontiers in Robotics and AI},
  volume={12},
  pages={1563923},
  year={2025},
  publisher={Frontiers Media SA}
}

@inproceedings{leite2016autonomous,
  title={Autonomous disengagement classification and repair in multiparty child-robot interaction},
  author={Leite, Iolanda and McCoy, Marissa and others},
  booktitle={2016 25th IEEE International Symposium on Robot and Human Interactive Communication (RO-MAN)},
  pages={525--532},
  year={2016},
  organization={IEEE}
}

@inproceedings{mutlu2009footing,
  title={Footing in human-robot conversations: how robots might shape participant roles using gaze cues},
  author={Mutlu, Bilge and Shiwa, Toshiyuki and others},
  booktitle={Proceedings of the 4th ACM/IEEE international conference on Human robot interaction},
  pages={61--68},
  year={2009}
}

@inproceedings{tennent2019micbot,
  title={Micbot: A peripheral robotic object to shape conversational dynamics and team performance},
  author={Tennent, Hamish and Shen, Solace and Jung, Malte},
  booktitle={2019 14th ACM/IEEE International Conference on Human-Robot Interaction (HRI)},
  pages={133--142},
  year={2019},
  organization={IEEE}
}

@inproceedings{shamekhi2019multimodal,
  title={A multimodal robot-driven meeting facilitation system for group decision-making sessions},
  author={Shamekhi, Ameneh and Bickmore, Timothy},
  booktitle={2019 international conference on multimodal interaction},
  pages={279--290},
  year={2019}
}

@inproceedings{pham2024embodied,
  title={Embodied mediation in group ideation--a gestural robot can facilitate consensus-building},
  author={Pham, Tuan Vu and Weisswange, Thomas H and Hassenzahl, Marc},
  booktitle={Proceedings of the 2024 ACM Designing Interactive Systems Conference},
  pages={2611--2632},
  year={2024}
}

@inproceedings{chen2025maintaining,
  title={Maintaining" Balanced" Conflict: Proactive Intervention Strategies of AI Voice Agents in Online Collaboration of Temporary Design Teams},
  author={Chen, XinHui and Yuan, Xiang and others},
  booktitle={Proceedings of the 2025 CHI Conference on Human Factors in Computing Systems},
  pages={1--19},
  year={2025}
}

@article{ochi2025robot,
  title={Robot-Mediated Multi-Party Conversation Aimed at Affect Improvement for Psychiatric Patients},
  author={Ochi, Keiko and Lala, Divesh and others},
  journal={IEEE Transactions on Affective Computing},
  year={2025},
  publisher={IEEE}
}

@article{correia2023robotic,
  title={Robotic gaze responsiveness in multiparty teamwork},
  author={Correia, Filipa and Campos, Joana and Melo, Francisco S and Paiva, Ana},
  journal={International Journal of Social Robotics},
  volume={15},
  number={1},
  pages={27--36},
  year={2023},
  publisher={Springer}
}

@inproceedings{jung2015using,
  title={Using robots to moderate team conflict: the case of repairing violations},
  author={Jung, Malte F and Martelaro, Nikolas and Hinds, Pamela J},
  booktitle={Proceedings of the tenth annual ACM/IEEE international conference on human-robot interaction},
  pages={229--236},
  year={2015}
}

@inproceedings{hoque2014empirical,
  title={An empirical robotic framework for interacting with multiple humans},
  author={Hoque, Mohammed Moshiul and Hossian, Quazi Delwar and others},
  booktitle={2013 International Conference on Electrical Information and Communication Technology (EICT)},
  pages={1--5},
  year={2014},
  organization={IEEE}
}

@inproceedings{gyoda2011mobile,
  title={Mobile care robot accepting requests through nonverbal interaction},
  author={Gyoda, Masahiko and Tabata, Tomoya and Kobayashi, Yoshinori and Kuno, Yoshinori},
  booktitle={2011 17th Korea-Japan Joint Workshop on Frontiers of Computer Vision (FCV)},
  pages={1--5},
  year={2011},
  organization={IEEE}
}

@article{chen2023outperformance,
  title={Outperformance of mall-receptionist android as inverse reinforcement learning is transitioned to reinforcement learning},
  author={Chen, Zhichao and Nakamura, Yutaka and Ishiguro, Hiroshi},
  journal={IEEE Robotics and Automation Letters},
  volume={8},
  number={6},
  pages={3350--3357},
  year={2023},
  publisher={IEEE}
}

@inproceedings{kobayashi2011assisted,
  title={Assisted-care robot dealing with multiple requests in multi-party settings},
  author={Kobayashi, Yoshinori and Gyoda, Masahiko and others},
  booktitle={Proceedings of the 6th international conference on Human-robot interaction},
  pages={167--168},
  year={2011}
}

@inproceedings{taylor2025rapidly,
  title={Rapidly built medical crash cart! lessons learned and impacts on high-stakes team collaboration in the emergency room},
  author={Taylor, Angelique and Tanjim, Tauhid and others},
  booktitle={2025 20th ACM/IEEE International Conference on Human-Robot Interaction (HRI)},
  pages={501--510},
  year={2025},
  organization={IEEE}
}

@inproceedings{gillet2022learning,
  title={Learning gaze behaviors for balancing participation in group human-robot interactions},
  author={Gillet, Sarah and Parreira, Maria Teresa and V{\'a}zquez, Marynel and Leite, Iolanda},
  booktitle={2022 17th ACM/IEEE International Conference on Human-Robot Interaction (HRI)},
  pages={265--274},
  year={2022},
  organization={IEEE}
}

@inproceedings{love2024towards,
  title={Towards explainable proactive robot interactions for groups of people in unstructured environments},
  author={Love, Tamlin and Andriella, Antonio and Aleny{\`a}, Guillem},
  booktitle={Companion of the 2024 ACM/IEEE International Conference on Human-Robot Interaction},
  pages={697--701},
  year={2024}
}

@inproceedings{bohus2010facilitating,
  title={Facilitating multiparty dialog with gaze, gesture, and speech},
  author={Bohus, Dan and Horvitz, Eric},
  booktitle={International Conference on Multimodal Interfaces and the Workshop on Machine Learning for Multimodal Interaction},
  pages={1--8},
  year={2010}
}

@inproceedings{joosse2014cultural,
  title={Cultural differences in how an engagement-seeking robot should approach a group of people},
  author={Joosse, Michiel P and Poppe, Ronald W and Lohse, Manja and Evers, Vanessa},
  booktitle={Proceedings of the 5th ACM international conference on Collaboration across boundaries: culture, distance \& technology},
  pages={121--130},
  year={2014}
}

@article{oertel2020engagement,
  title={Engagement in human-agent interaction: An overview},
  author={Oertel, Catharine and Castellano, Ginevra and others},
  journal={Frontiers in Robotics and AI},
  volume={7},
  pages={92},
  year={2020},
  publisher={Frontiers Media SA}
}

@article{pillet2010making,
  title={Making way and making sense: Including newcomers in interaction},
  author={Pillet-Shore, Danielle},
  journal={Social Psychology Quarterly},
  volume={73},
  number={2},
  pages={152--175},
  year={2010},
  publisher={Sage Publications Sage CA: Los Angeles, CA}
}

@article{traeger2020vulnerable,
  title={Vulnerable robots positively shape human conversational dynamics in a human--robot team},
  author={Traeger, Margaret L and Strohkorb Sebo, Sarah and others},
  journal={Proceedings of the National Academy of Sciences},
  volume={117},
  number={12},
  pages={6370--6375},
  year={2020},
  publisher={National Academy of Sciences}
}

@inproceedings{gillet2021robot,
  title={Robot gaze can mediate participation imbalance in groups with different skill levels},
  author={Gillet, Sarah and Cumbal, Ronald and Pereira, Andr{\'e} and Lopes, Jos{\'e} and Engwall, Olov and Leite, Iolanda},
  booktitle={Proceedings of the 2021 ACM/IEEE International Conference on Human-Robot Interaction},
  pages={303--311},
  year={2021}
}

@inproceedings{mongile2023if,
  title={What if a social robot excluded you? using a conversational game to study social exclusion in teen-robot mixed groups},
  author={Mongile, Sara and Pusceddu, Giulia and others},
  booktitle={Companion of the 2023 ACM/IEEE International Conference on Human-Robot Interaction},
  pages={208--212},
  year={2023}
}

@article{matsuyama2015four,
  title={Four-participant group conversation: A facilitation robot controlling engagement density as the fourth participant},
  author={Matsuyama, Yoichi and Akiba, Iwao and Fujie, Shinya and Kobayashi, Tetsunori},
  journal={Computer Speech \& Language},
  volume={33},
  number={1},
  pages={1--24},
  year={2015},
  publisher={Elsevier}
}

@inproceedings{hedayati2022predicting,
  title={Predicting positions of people in human-robot conversational groups},
  author={Hedayati, Hooman and Szafir, Daniel},
  booktitle={2022 17th ACM/IEEE International Conference on Human-Robot Interaction (HRI)},
  pages={402--411},
  year={2022},
  organization={IEEE}
}

@inproceedings{hedayati2020reform,
  title={Reform: Recognizing f-formations for social robots},
  author={Hedayati, Hooman and Muehlbradt, Annika and Szafir, Daniel J and Andrist, Sean},
  booktitle={2020 IEEE/RSJ International Conference on Intelligent Robots and Systems (IROS)},
  pages={11181--11188},
  year={2020},
  organization={IEEE}
}

@inproceedings{mok2016effects,
  title={Effects of proactivity and expressivity on collaboration with interactive robotic drawers},
  author={Mok, Brian},
  booktitle={2016 11th ACM/IEEE International Conference on Human-Robot Interaction (HRI)},
  pages={633--634},
  year={2016},
  organization={IEEE}
}

@article{rozo2016learning,
  title={Learning controllers for reactive and proactive behaviors in human--robot collaboration},
  author={Rozo, Leonel and Silv{\'e}rio, Joao and Calinon, Sylvain and Caldwell, Darwin G},
  journal={Frontiers in Robotics and AI},
  volume={3},
  pages={30},
  year={2016},
  publisher={Frontiers Media SA}
}

@inproceedings{tanjim2025help,
  title={Help or Hindrance: Understanding the Impact of Robot Communication in Action Teams},
  author={Tanjim, Tauhid and George, Jonathan St and Ching, Kevin and Taylor, Angelique},
  booktitle={2025 34th IEEE International Conference on Robot and Human Interactive Communication (RO-MAN)},
  pages={1460--1465},
  year={2025},
  organization={IEEE}
}

@inproceedings{tanjim2025human,
  title={Human-Robot Teaming Field Deployments: A Comparison Between Verbal and Non-verbal Communication},
  author={Tanjim, Tauhid and Ekpo, Promise and others},
  booktitle={2025 34th IEEE International Conference on Robot and Human Interactive Communication (RO-MAN)},
  pages={1699--1704},
  year={2025},
  organization={IEEE}
}

@article{kanda2010communication,
  title={A communication robot in a shopping mall},
  author={Kanda, Takayuki and Shiomi, Masahiro and others},
  journal={IEEE Transactions on Robotics},
  volume={26},
  number={5},
  pages={897--913},
  year={2010},
  publisher={IEEE}
}

@inproceedings{taylor2024towards,
  title={Towards collaborative crash cart robots that support clinical teamwork},
  author={Taylor, Angelique and Tanjim, Tauhid and Cao, Huajie and Lee, Hee Rin},
  booktitle={Proceedings of the 2024 ACM/IEEE International Conference on Human-Robot Interaction},
  pages={715--724},
  year={2024}
}

\end{document}